\documentstyle[12pt]{article}

\def\beq{\begin{equation}}
\def\eeq{\end{equation}}
\def\bea{\begin{eqnarray}}
\def\eea{\end{eqnarray}}

\def\ba{\begin{array}}
\def\ea{\end{array}}

\def\,{\"{U}}
\def\6{\.{I}}

\begin{document}

\title{PT-symmetric Solutions of Schr\"{o}dinger Equation with position-dependent
mass via Point Canonical Transformation}
\author{Cevdet Tezcan\\
Faculty of Engineering, Ba\c{s}kent University, Bagl{\i}ca Campus, Ankara, Turkey\\[.5cm]
 Ramazan Sever$\thanks{Corresponding author:
sever@metu.edu.tr}$\\
Department of Physics, Middle East Technical University, 06531
Ankara, Turkey }

\date{\today}
\maketitle

\begin{abstract}
PT-symmetric solutions of Schr\"{o}dinger equation are obtained for
the Scarf and generalized harmonic oscillator potentials with the
position-dependent mass. A general point canonical transformation is
applied by using a free parameter. Three different forms of mass
distributions are used. A set of the energy eigenvalues of the bound
states and corresponding wave functions for
target potentials are obtained as a function of the free parameter.
\end{abstract}

\baselineskip 0.9cm

\newpage

\section{Introduction}

Exact solutions of the effective mass Schr\"{o}dinger equation(SE)
for some physical potentials have received much attention. Important
applications are obtained in the fields of material science and
condensed matter physics such as semiconductors [1], quantum well
and quantum dots [2], $^{3}H$ clusters [3], quantum liquids [4],
graded alloys and semiconductor heterostructures [5,6]. Recently,
number of exact solutions on these topics increased [6-23]. Various
solution methods are used in the calculations. The point canonical
transformations (PCT) is one of these methods providing exact
solution of energy eigenvalues and corresponding eigenfunctions
[24-28]. It is also used for solving the Schr\"{o}dinger equation
with position-dependent effective mass for some potentials [8-13].
In the present work, we solve three different potentials with the
three mass distributions. The point canonical transformation is
taken in the more general form introducing a free parameter. This
general form of the transformation will provide us a set of
solutions for different values of the free parameter.

On the other hand, there has been considerable work on non-Hermitian
Hamiltomians in recent years. Much attentioan has also been focused
on ${\cal PT}$-symmetric Hamiltonians. Following the early studies
of Bender {\it et al}. [29], the ${\cal PT}$-symmetry formulation
has been successfully utilized by many authors [30-36]. The ${\cal
PT}$-symmetric but non-Hermitian Hamiltonians have real spectra
whether the Hamiltonians are Hermitian or not. Non-Hermitian
Hamiltonians with real or complex spectra have also been
analyzed by using different methods [31-34,37]. Non-Hermitian but ${\cal PT}$%
-symmetric models have applications in different fields, such as
optics [38], nuclear physics [39], condensed matter [40] and
population biology [41]. There are some recent works on these
topics[42-44].

The contents of the paper is as follows. In section 2, we present
briefly the solution of the Schr\"{o}dinger equation by using point
canonical transformation. In section 3, we introduce some
applications for specific potentials. Results are discussed in
section 4.

\section{Method}
We write the most general form of the Hamiltonian as [45]

\begin{equation}
H=\frac{1}{4}\left[ m^{\eta}p m^{\lambda} p m^{\mu}+m^{\mu}p
m^{\lambda} p m^{\eta}+ V(x)\right]
\end{equation}
where the parameters $\eta, \lambda$ and $\mu$ are called the
ambiguity parameters. They are constraint by the relation $\eta
+\lambda+\mu=-1$. Different forms of the Hamiltonian are used in the
literature depending on the choices of set of parameters. Here, we
take the set such that $\eta=\mu=0$, $\lambda=1$ [46,47] introduced
by Ben Danie-Duke. Thus, the Hamiltonian is invariant under
instantaneous Galilean transformation [47]. The Schr\"{o}dinger
equation with $\hbar = 1$ takes the form

\begin{equation}
-\frac{d}{dx}\left[\frac{1}{2m(x)}\frac{d\Psi}{dx}\right]+
V(x)\Psi(x)=E\Psi(x).
\end{equation}
The wave function should be continuous at the mass discontinuity and
verify the following condition

\begin{equation}
\left.\frac{1}{m(x)}\frac{d\Psi(x)}{dx}\right|_{-}=\left.\frac{1}{m(x)}\frac{d\Psi(x)}{dx}\right|_{+}.
\end{equation}
On the other hand a Hamiltonian is said to be PT-symmetric if

\begin{equation}
\left[PT,H\right]=0
\end{equation}
where $P$ is the parity operator and $T$ is the time reversal
operator. They act on the position and momentum states as

\begin{equation}
P: x\rightarrow -x, \hspace{2 mm}P\rightarrow -P \indent and \indent
T: x\rightarrow x, \hspace{2 mm}P\rightarrow -P, \indent
i\rightarrow -i.
\end{equation}
Thus, we get the following conditions to have a PT-symmetric
Hamiltonian

\begin{equation}
m(x)=m(-x)\indent and \indent V^{*}(-x)=V(x).
\end{equation}
We shall use a general form of PCT with a free parameter to solve
the Schr\"{o}dinger equation for any potential $V(x)$. Defining the
transformation with a free parameter $\beta$

\begin{equation}
\Psi(x)=m^{\beta}(x)\phi(x)
\end{equation}
The SE takes

\begin{equation}
-\frac{1}{2m}\left[\phi^{''}+\left(2\beta-1\right)\frac{m^{'}}{m}\phi^{'}+\beta
\left(\beta-2\right)\left(\frac{m^{'}}{m}\right)^{2}\phi+\beta
\left(\frac{m^{''}}{m}\right)\phi\right]+ V(x)\phi=E\phi.
\end{equation}
It is solved by Roy [42] for $\beta=\frac{1}{4}$. Roy obtained this
form by making the transformation

\begin{equation}
y=\int^x m(t)^{\frac{1}{2}}dt.
\end{equation}

In the computations, three different position-dependent mass
distributions[47] will be used. The reference potentials are
PT-symmetric Scarf II[49,50] and generalized generalized harmonic
oscillator[51] potentials. We will consider two different values of
$\beta$.

\section{Case A:$\beta=\frac{1}{2}$}

For $\beta=\frac{1}{2}$, $Eq.(8)$ has the following  compact form

\begin{equation}
-\frac{1}{2}\frac{d^{2}\phi}{dx^{2}}+\Omega(x)\phi(x)=E\phi(x)
\end{equation}
where

\begin{equation}
\Omega(x)=\frac{3}{8}\left(\frac{m^{'}}{m}\right)^{2}-\frac{1}{4}\left(\frac{m^{''}}{m}\right)+mV-\left(m-1\right)E
\end{equation}
Eq.(10) has the same form with the constant mass SE.

\subsection{Mass Distribution $m(x)=\left(\frac{\alpha+x^{2}}{1+x^{2}}\right)^{2}$}

i) The Scarf II potential is

\begin{equation}
\Omega(x)=-\lambda sech^{2}(x)-i\mu sech(x)\tanh(x)
\end{equation}
The energy eigenvalues and corresponding wave functions are

\begin{equation}
E_{n}=-\left(n-p-1\right)^{2}, \hspace{15 mm} n=0, 1, 2,
...<\frac{s+t-1}{2}
\end{equation}
and

\begin{equation}
\phi_{n}(x)=\frac{\Gamma(n-2p+\frac{1}{4})}{n!\hspace{1
mm}\Gamma(\frac{1}{2}-2p)}\hspace{1 mm}z^{-p}\hspace{1
mm}(z^{*})^{-q}\hspace{2
mm}P_{n}^{(-2p-\frac{1}{2},-2q-\frac{1}{2})}\left(i\sinh(y)\right)
\end{equation}
where

\begin{equation}
z=\frac{1-i\sinh(x)}{2},
\end{equation}

\vspace{10 mm}

\hspace{62 mm}
$p=-\frac{1}{4}\pm\frac{1}{2}\sqrt{\frac{1}{4}+\lambda+\mu}$,

\begin{equation}
=-\frac{1}{4}\pm\frac{t}{2},
\end{equation}

\vspace{10 mm}

\hspace{62 mm}
$q=-\frac{1}{4}\pm\frac{1}{2}\sqrt{\frac{1}{4}+\lambda-\mu}$

\begin{equation}
=-\frac{1}{4}\pm\frac{s}{2}.
\end{equation}
For the Scarf II potential and the position-dependent mass case, we
obtain the effective potential as

\begin{equation}
V(x)= \left(\frac{1+x^{2}}{\alpha+x^{2}}\right)^{2}\left\{
\Omega(x)-
4(1-\alpha)^{2}\frac{x^{2}}{(1+x^{2})^{2})(\alpha+x^{2})^{2}}+
\frac{(1-\alpha)(1-3x^{2})} {(1+x^{2})^{2}(\alpha+x^{2})}+
\left[(\frac{\alpha+x^{2}}{1+x^{2}})^{2}-1\right]E \right\}
\end{equation}
where

\begin{equation}
\frac{m^{'}}{m}=4(1-\alpha)\frac{x}{(1+x^{2})(\alpha+x^{2})}
\end{equation}
and

\begin{equation}
\frac{m^{"}}{m}=\frac{4(1-\alpha)}{(1+x^{2})^{2}(\alpha +x^{2})^{2}}
\left[2(1- \alpha x^{2})+(\alpha + x^{2})(1-3x^{2})\right].
\end{equation}

ii) PT-symmetric generalized harmonic oscillator

\begin{equation}
\Omega(x)=(x-i\varepsilon)^{2}+\frac{g^{2}-\frac{1}{4}}{(x-i\varepsilon)^{2}}.
\end{equation}
The energy eigenvalues and corresponding wave functions are

\begin{equation}
E_{n}=4n-2qg+2, \hspace{15 mm} n=0, 1, 2, ...
\end{equation}
and

\begin{equation}
\phi_{n}(x)=e^{-\frac{1}{2}(x-i\varepsilon)^{2}}(x-i\varepsilon)^{-pg+\frac{1}{2}}L_{n}^{-qg}\left((x-i\varepsilon\right)^{2})
\end{equation}
where $q=\pm1$ is called quasi-parity. Here the effective potential
is

\begin{eqnarray}
V(x)=\left(\frac{1+x^{2}}{\alpha+x^{2}}\right)^{2}\left\{\Omega(x)-\frac{3}{8}\left(\frac{m^{'}}{m}\right)^{2}+\frac{1}{4}\left(\frac{m^{''}}{m}\right)+\left[\left(\frac{\alpha+x^{2}}{1+x^{2}}\right)^{2}-1\right]E\right\}
\end{eqnarray}
where $\frac{m^{'}}{m}$ and $\frac{m^{''}}{m}$ are given in
$Eqs.(20,21)$.

\subsection{Mass Distribution $M(x)=\left(\frac{\alpha+x^{2}}{1+x^{2}}\right)^{4}=m^{2}$}

i)Scarf II potential

The form of the Scarf II  potential is given in Eq. (13).

where $\frac{m^{'}}{m}$, $\frac{m^{''}}{m}$ and $\Omega(x)$  are
given in $Eqs.(20,21, 13)$. Solution of the SE for the Scarf II
potential gives us the energy eigenvalues and corresponding wave
functions as

\begin{equation}
E=-(n-p-1)^{2} \hspace{15 mm} n=0, 1, 2,...<\frac{s+t-1}{2}
\end{equation}

ii) PT-symmetric generalized oscillator

It is given in $Eq. (22)$

The potential is given in $Eq. (13)$

\begin{eqnarray}
V(x)=\left(\frac{1+x^{2}}{\alpha+x^{2}}\right)^{4}
\{\Omega(x)-(\frac{m^{'}}{m})^{2}+
\frac{1}{2}(\frac{m^{''}}{m})+[(\frac{\alpha+x^{2}}{1+x^{2}})^{2}-1]E\}
\end{eqnarray}

\begin{eqnarray}
V(x)=\left(\frac{1+x^{2}}{\alpha+x^{2}}\right)^{4}\left\{\Omega(x)-\left(\frac{m^{'}}{m}\right)^{2}+
\frac{1}{2}\left(\frac{m^{''}}{m}\right)+\left[\left(\frac{\alpha+x^{2}}{1+x^{2}}\right)^{2}-1\right]E\right\}
\end{eqnarray}
where $\frac{m^{'}}{m}$, $\frac{m^{''}}{m}$ and $\Omega(x)$  are
given in $Eqs.(20,21, 22)$. The wave function is

\begin{equation}
\phi_{n}(y)=\frac{\Gamma(n-2p+\frac{1}{4})}{n!\hspace{1
mm}\Gamma(\frac{1}{2}-2p)}\hspace{1 mm}z^{-p}\hspace{1
mm}(z^{*})^{-q}\hspace{2
mm}P_{n}^{(-2p-\frac{1}{2},-2q-\frac{1}{2})}\left(i\sinh(y)\right)
\end{equation}
where z, p and q are defined in $Eqs. (16, 17, 18)$.

\section{Case B:$\beta=\frac{2-\gamma}{4}$}

Here, we define a new independent variable

\begin{equation}
y=\int^x m^{\frac{\gamma}{2}}(t)dt.
\end{equation}
Then, $Eq.(3)$ takes the form

\begin{equation}
-\frac{1}{2m}\left[m^{\gamma}\phi^{''}+\left(\frac{\gamma}{2}+
\left(\beta-1\right)m^{\frac{\gamma}{2}-1}m^{'}\right)\phi^{'}+
\beta\left(\beta-1\right)\left(\frac{m^{'}}{m}\right)^{2}\phi+\beta\left(\frac{m^{''}}{m}\right)\phi\right]+
V\phi=E\phi
\end{equation}
To remove the term involving first derivative of the wave function,
we impose

\begin{center}
$\frac{\gamma}{2}+2\beta-1=0$.
\end{center}
This is the constraint on the parameter $\beta$ to get the exact
solution.

Thus, we get

\begin{equation}
-\frac{1}{2}\phi^{''}+\Omega(y)\phi=E\phi
\end{equation}
where

\begin{equation}
\Omega(y)=-\frac{\beta}{2}m^{-\gamma}\left[\left(\beta-2\right)\left(\frac{m^{'}}{m}\right)^{2}+\frac{m^{''}}{m}\right]+\left(V-E\right)m^{1-\gamma}+E
\end{equation}
and also

\begin{equation}
V(x)=m^{\gamma-1}\left[\Omega(y)+\frac{2-\gamma}{\gamma
m^\gamma}\left[-\frac{\gamma+6}{4}\left(\frac{m^{'}}{m}\right)^{2}+\frac{m^{''}}{m}\right]+E\left(1-m^{1-\gamma}\right)\right].
\end{equation}

\subsection{Mass Distribution $m(x)=\left(\frac{\alpha+x^{2}}{1+x^{2}}\right)^{\frac{2}{\gamma}}$}

The new independent variable is

\begin{equation}
y=\left[x+(\alpha-1)\tan^{-1}(x)\right].
\end{equation}

i) Scarf II potential

The scarf II potential for the potential and position dependent mass
has the form

\begin{equation}
V(x)=m^{\gamma-1}\left[\Omega(y)+\frac{2-\gamma}{8m^\gamma}\left[-\frac{\gamma+6}{4}\left(\frac{m^{'}}{m}\right)^{2}+\frac{m^{''}}{m}\right]+E\left(1-m^{1-\gamma}\right)\right]
\end{equation}
where $\frac{m^{\prime}}{m}$ and $\frac{m^{\prime\prime}}{m}$ are
given in $Eqs. (35,36)$.

\begin{equation}
\frac{m^{'}}{m}=\frac{4(1-\alpha)}{\gamma}\frac{x}{(1+x^{2})(\alpha+x^{2})}
\end{equation}
and

\begin{equation}
\frac{m^{''}}{m}=\frac{4{(1-\alpha}}{\gamma (1+x^{2})^{2}\hspace{1
mm}(\alpha+x^{2})^{2}}\left
[2(k-1)(1-\alpha)x^{2}+(\alpha+x^{2})(1-3x^{2})\right]
\end{equation}
where $k=\frac{2}{\gamma}$

ii) PT-symmetric generalized oscillator

Solution of the SE for the PT-symmetric generalized oscillator, $Eq.
(22)$, gives us energy eigenvalues and corresponding wave functions
as

\begin{equation}
E_{n}=4n-2q\rho+2, \hspace{15 mm}  n=0, 1, 2,...
\end{equation}
and

\begin{equation}
\phi_{n}(y)=e^{-\frac{1}{2}(y-i\varepsilon)^{2}}(y-i\varepsilon)^{-p\rho+\frac{1}{2}}\hspace{2
mm}L_{n}^{-q\rho}\left((y-i\varepsilon)^{2}\right)
\end{equation}
where $q=\pm1$ is called quasi-parity. For the potential and
position dependent mass, we get

\section{Conclusions}
We have applied the point canonical transformation in a general form
by introducing a free parameter to solve the Schr\"{o}dinger
equation for the Rosen-Morse and Scarf potentials with spatially
dependent mass. We have obtained a set of exactly solvable target
potentials by  using two position-dependent mass distributions.
Energy eigenvalues and corresponding wave functions for the target
potentials are written in the compact form.

\section{Acknowledgements}

This research was partially supported by the Scientific and
Technological Research Council of Turkey.

\end{document}